\documentclass[aps,prd,reprint,groupedaddress]{revtex4-1}
\usepackage{hyperref}
\usepackage{graphicx}
\usepackage{color}
\usepackage{amsmath}
\usepackage{amssymb}
\hypersetup{
    pdfnewwindow=true,      
    colorlinks=true,       
    linkcolor=blue,          
    citecolor=blue,        
    filecolor=blue,      
    urlcolor=blue           
}

\newcommand{\lrr}{Living Rev. Relativ.}
\newcommand{\mnras}{MNRAS}
\newcommand{\apjl}{ApJ Lett.}
\newcommand{\apjs}{ApJS}

\newcommand{\araa}{Annu. Rev. Astron. Astrophys.}

\begin{document}

\title{Observational consequences of structured jets from \\ neutron star mergers in the local Universe}

\author{Nihar Gupte}
\affiliation{Department of Physics, University of Florida, Gainesville, FL 32611, USA}
\author{Imre Bartos}
\email{imrebartos@ufl.edu}
\affiliation{Department of Physics, University of Florida, Gainesville, FL 32611, USA}

\begin{abstract}
The recent multi-messenger discovery of binary neutron star (BNS) merger GW170817 showed that $\gamma$-ray emission in short GRBs is wider than the central energetic narrow cone, and weakly expands out to tens of degrees. Here we explore some of the observational consequences of this structured emission, taking the reconstructed angular emission profile of gamma-ray burst GRB170817A to be typical. We calculate the expected fraction of gravitational-wave observations from BNS mergers that will have an observed GRB counterpart to be $\sim30\%$, implying that joint gravitational-wave and GRB observations will be common. Further, we find that $\sim10\%$ of observed short GRBs occur within 200\,Mpc. Finally, we estimate a BNS merger rate of $\sim500$\,Gpc$^{-3}$yr$^{-1}$.
\end{abstract}

\maketitle

\section{Introduction}

The observation of binary neutron star (BNS) merger GW170817 demonstrated the impact of combining information from multiple astrophysical messengers \cite{2017PhRvL.119p1101A,2017ApJ...848L..12A,2017ApJ...848L..13A,2017ApJ...850L..35A}. Multi-messenger information from this one event, among others, firmly established the connection between neutron star mergers and short gamma-ray bursts (GRBs), gave us information on the structure of the relativistic outflow \cite{2018arXiv180502870A,2017arXiv171203237L,2018Natur.554..207M,2018PTEP.2018d3E02I,2017arXiv171005896G,2017ApJ...848L..25H,2018arXiv180502870A,2018arXiv180207328V} and composition of the dynamical ejecta \cite{2017ApJ...851L..21V,2017Sci...358.1565E,2017arXiv171109638W}, constrained the maximum mass and equation of state of neutron stars \cite{2017ApJ...850L..19M,2018ApJ...852L..29R}, and gave an independent estimate of the Hubble constant \cite{2017Natur.551...85A}.

The detection of a GRB from the first BNS merger observed through GWs was unexpected \cite{2012ApJ...746...48M,2015MNRAS.448.3026W}. Gamma-ray emission from short GRBs is beamed, reducing the rate of events observable from Earth. The level of beaming can be estimated using the sudden steepening of GRB afterglow light curves (the so-called jet break) recorded for some GRBs. This yields a typical beaming half angle of $\sim10^\circ$ \cite{2014ARA&A..52...43B}. This means that only one in every $\sim100$ GRBs should be observable \cite{2012ApJ...746...48M}. Similar beaming can be inferred from comparing the observed rate of short GRBs to the expected rate of BNS mergers, the latter of which can be estimated using the observed Galactic binary population and population synthesis models \cite{2004ApJ...601L.179K,2010CQGra..27q3001A,2014ARA&A..52...43B}.

GWs, on the other hand, are emitted by BNS mergers in all directions. The difference between the GW amplitude along the direction of strongest emission--the orbital axis, and the direction-averaged emission is only a factor of $1.5$ \cite{2011CQGra..28l5023S}. This means that GWs can be observed with limited dependence on the source orientation. Therefore, for highly beamed gamma emission we expect a high fraction of GW detections not to be accompanied by a detected GRB. Assuming a GRB beaming factor of 100, accounting for the fact that gamma-rays will be emitted along the orbital axis of the binary after merger, and assuming that all GRBs facing towards Earth can be detected within the distance range of GW detectors, about $1.5^3/100\approx3\%$ of GW detections should have an observable GRB counterpart.

This paradigm changed with the discovery of GWs from the BNS merger GW170817 by the LIGO and Virgo GW detectors, which was accompanied by a GRB, GRB170817A, discovered by the Fermi Gamma-ray Burst Monitor (Fermi-GBM) and the Anti-Coincidence Shield for the Spectrometer for the International Gamma-Ray Astrophysics Laboratory (INTEGRAL) \cite{2017ApJ...848L..13A}. The coincidence for the first discovery was previously deemed unlikely. Further, from the identification of the host galaxy together with the GW signal one could estimate the binary's inclination, found to be $32^{+10}_{-13}$\,deg \cite{2018arXiv180404179F}. Short GRBs as a group cannot be detected up to this inclination. A jet opening half angle $\theta_{\rm j} = 30^\circ$ would imply a beaming factor of $f_{\rm b}=(1-\cos(\theta_{\rm j})^{-1} \approx 7$. This is inconsistent with the (non-collapsar) short-GRB rate density of $\sim10$\,Gpc$^{-3}$yr$^{-1}$ \cite{2006ApJ...650..281N} and the binary neutron star merger rate density of $\sim 10^{3}$\,Gpc${-3}$yr$^{-1}$ \cite{2010CQGra..27q3001A,2017PhRvL.119p1101A}. 

We do not typically detect GRBs with these inclinations at cosmological distances. Other than the possibility that GRB170817A is unusual, this can be explained if we assume that gamma-ray emission weakens at large observing angles $\theta_{\rm obs}$ measured from the GRB jet axis. At low $\theta_{\rm obs}$ gamma-ray luminosity is high, and GRBs can be detected from large distances. At greater $\theta_{\rm obs}$, gamma-ray luminosity diminishes and only the closest events can be detected. In this scenario the rate density of observed GRBs is determined by the narrow cone of high-luminosity GRB which has a large effective beaming factor. This picture is further corroborated by the measured isotropic-equivalent energy of GRB170817A, $E_{\rm iso}\approx 3\times10^{46}$\,erg \cite{2017ApJ...848L..13A}. This is about a factor of a 1000 below the isotropic-equivalent energy of the weakest GRB previously observed with known redshift \cite{2014ARA&A..52...43B}. 

Observations of the GRB afterglow provided a wealth of additional information on the structure of the outflow that produce gamma rays. The delayed onset of the X-ray afterglow is consistent with $\theta_{\rm obs}$ being greater than the opening half angle, i.e. that the GRB was observed off-axis \cite{2017Natur.551...71T,2017ApJ...848L..25H}. 

The afterglow's temporal and spectral evolution can also be used to reconstruct the properties of the relativistic outflow from the source. Here, an interesting result is that the relativistic outflow interacts with the lower-velocity ejecta from the merger, affecting both the afterglow and the gamma-ray emission. 

Numerical simulations of the process show that the observed afterglow is inconsistent with the simple {\it top-hat} jet, in which the relativistic outflow produces uniform emission within the opening angle and zero emission outside of it. Observations are, however, consistent with structured relativistic outflow, more specifically the off-axis observation of a narrow cone of ultra-relativistic material surrounded by a slower outflow that extends to greater angles \cite{2017arXiv171203237L,2018ApJ...856L..18M}. Another possible explanation is a quasi-spherical, mildly relativistic ejecta produced by the energy injection of a narrow relativistic jet into slow-moving ejecta \cite{2017Sci...358.1559K,2018Natur.554..207M,2018ApJ...858L..15D,2018arXiv180609693M}. Observations cannot yet differentiate between these latter scenarios.

For either of these possibilities, gamma-ray emission, at least for GRB170817A, is more directionally extended than previous estimates of GRB beaming suggested. This is good news for the joint observability of GWs and GRBs from BNS mergers, and possibly neutron star and black hole mergers. While the weak extended emission does not affect the detection rate of GRBs at cosmological distances, within the limited distance range of GW observations it can play an important role.

In this paper we investigate the observability and inferred rate of BNS mergers in light of structured gamma-ray emission. For this we assume that GRB170817A is a typical short GRB, and adopt the reconstructed angular gamma-ray emission profile of a structured outflow computed by Margutti {\it et al.} \cite{2018ApJ...856L..18M}. Using this profile, we examine: what fraction of BNS mergers detected via GWs that will be also detected via GRBs; whether there could be a significant population of nearby detectable GRBs, within $\sim200$\,Mpc; and the inferred rate of BNS mergers.

\section{Method}
\label{sec:method}

We used Monte Carlo simulations to determine the detectability of GWs and GRBs from BNS mergers. For each realization, we placed a merger in space at a random location assuming homogeneous distribution. We selected a random direction for the binary orbital axis, which we assumed to be the same as the GRB jet axis. For each event, we separately determined the detectability of the GW signal using Advanced LIGO and Advanced Virgo, and the GRB emission using Fermi-GBM and Swift-BAT. 

\subsection{Gravitational Waves}

To determine whether a GW can be detected, we required that its expected signal-to-noise ratio (SNR) in the network of GW detectors, $\rho_{\rm network}$, exceeds a threshold value of 12. We chose this threshold as it corresponds to a false alarm rate of $\sim 10^{-2}$\,yr$^{-1}$ \cite{AbbottProspects2018}. 

The GW SNR $\rho$ for a given detector is determined by the detector's sensitivity, as well as the binary's inclination and orientation compared to the detector. To characterize the sensitivity of a given detector to BNS mergers, we define its horizon distance $D_{\rm h}$ as follows. An optimally oriented merger in the optimal direction compared to the detector at $D_{\rm h}$ distance produces a GW signal in the detector with an expected SNR of 8. 

For a source located at distance $r$, polar angle $\theta_{\rm gw}$ and azimuthal angle $\phi_{\rm gw}$ relative to the detector's axes (see Fig. 1 in \cite{schutz2011networks}), its mean power SNR over the ensemble (denoted by $\langle \rangle$) for a single detector can be written as
\begin{equation} \label{SNR}
\langle \rho^2 \rangle = P(\theta, \phi) \frac{D^{2}_{\rm h}}{r^2}.
\end{equation}
Here, $P(\theta, \phi)$ is the \textit{antenna power pattern} of a single interferometer \cite{schutz2011networks}.

To find the network SNR, we combine the SNR of individual detectors:
\begin{equation}
\rho_{\rm network} \equiv \sqrt{\sum_{\rm i} \langle \rho^{2}_{\rm i} \rangle},
\end{equation}
where the sum is over the GW detectors in the network.

We examine multiple observation scenarios by selecting different horizon distances for the detectors. We consider LIGO/Virgo's next observing period, O3, by adopting the expected sensitivities for the ''Late" phase in Table 1 of \cite{AbbottProspects2018}. Note that these values are ranges, which need to be multiplied by 2.26 to obtain the horizon distance (e.g., \cite{2013CQGra..30l3001B}). Accounting for this, we adopt a late-phase BNS merger horizon distance range of $270-380$\,Mpc for LIGO and $150-260$\,Mpc for Virgo. We further consider the LIGO/Virgo network at its design sensitivity, using the ''Design" phase from the same table used in \cite{AbbottProspects2018}, following the same method. This gives a design phase BNS horizon distance of $430$\,Mpc for LIGO and 280\,Mpc for Virgo. Finally, for a crude comparison, we considered a future 3rd-generation GW observatory whose horizon distance is $10$ times that of LIGO's at design sensitivity. For this case, we assume that only one of such detectors is operational, and for simplicity do not account for cosmological effects or the evolution of merger rate densities. This gives a BNS horizon distance of 4300\,Mpc.

\subsection{Gamma-Ray Bursts}

\begin{figure}
\begin{center}
\includegraphics[width=1.08\linewidth]{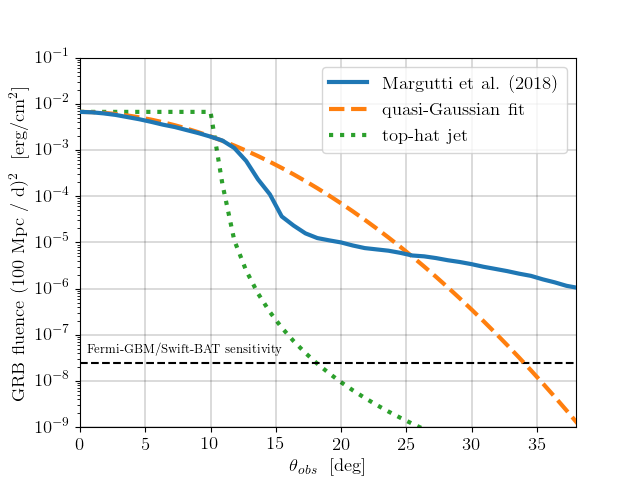}
\end{center}
\caption{Expected fluence of a short GRB similar to GRB170817A for a source at 100\,Mpc from Earth, as a function of viewing angle. Our fiducial model is the simulated structured jet of Margutti {\it et al.} (blue). For comparison, we show a quasi-Gaussian best fit to the fiducial model (green), as well as the expected profile of a top-hat jet with $10^\circ$ beaming half angle, and off-axis emission due to relativistic effects assuming a Lorentz factor $\Gamma=100$. We also show the approximate detection threshold of Fermi-GBM and Swift-BAT.}
\label{fig:fluence}
\end{figure}

We adopted the structured jet model of Margutti {\it et al.} \cite{2018ApJ...856L..18M}, which they obtained using numerical simulations fit to reproduce the afterglow observations of GRB170817A. The expected fluence at Earth as a function of $\theta_{\rm obs}$ for this model is shown 
in Fig. \ref{fig:fluence}, adopted from Margutti {\it et al.} \cite{2018ApJ...856L..18M} (their Fig. 4). For comparison, we also show a quasi-Gaussian best fit on the simulated profile, $E_{\rm fit} = E_0 e^{-(\theta_{obs}/\theta_c)^\alpha}$, with $E_0=10^{52}$\,erg, $\alpha = 1.9$ and $\theta_c = 9^\circ$ \cite{2018ApJ...856L..18M}. As another comparison, we show a simple top-hat jet profile. In the top-hat model, beyond uniform on-axis emission within the jet opening half angle $\theta_{\rm, j}$, we additionally account for the fact that gamma-ray emission can also be observed off-axis due to relativistic effects. This emission is Doppler shifted and weakened compared to the on-axis emission. Here, we adopt the off-axis emission model of \cite{2018PTEP.2018d3E02I} (see their Appendix A), with $\theta_{\rm j} = 10^\circ$ and Lorentz factor $\Gamma=100$.

For the results below, we get similar values if we adopt the quasi-Gaussian fit or the structured jet model. Below we only show results for the structured jet model.

We make the simplifying assumption that the GRB spectrum, which is different at different angles, has no further effect on detectability. We fully describe detectability with the observed gamma-ray flux.

For each Monte Carlo realization, we use the structured jet model to determine whether a GRB would be detected. We take into account both the Fermi-GBM and Swift-BAT detectors. For both Fermi-GRB and Swift-BAT, we consider a source to be detectable if its fluence exceeds $2.5 \times 10^{-8}$ erg\,cm$^{-2}$ \cite{2016ApJ...818..110B}. Fermi-GBM monitor about $70\%$ of the sky, while Swift-BAT monitors about 15\%. Therefore we assign a probability of $0.7+(1-0.7)\cdot0.15 = 0.745$ of detection to each source that exceeds the $2.5\times 10^{-8}$ erg\,cm$^{-2}$ fluence threshold.

\section{Results}

\subsection{What fraction of GW detections will be accompanied by an observed GRB?}

We first calculated the volume (hereafter {\it detection volume}) within which a GW signal from a BNS merger could be detected for a late-phase and design-phase LIGO/Virgo network \cite{AbbottProspects2018}, as well as for a hypothetical 3rd-generation detector. This scenario only uses information from GWs. Results are shown in Table \ref{table:gw-only}.

\begin{table}
	\centering
	\begin{tabular}{|c|c|c|}
		\hline
		GW Phase       & GW-only [Gpc$^3$] & GW+GRB [Gpc$^{3}$]  \\ \hline
		Late phase     & $0.04-0.05$       & $0.013-0.019$       \\ \hline
		Design phase   & $0.06$            & 0.023               \\ \hline
		3rd Gen.       & $0.34$            & 0.08                \\ \hline
	\end{tabular}
\label{table:gw-only}    
\caption{Expected volumes within which a GW signal from a BNS merger could be detected for LIGO/Virgo late phase, LIGO/Virgo at design sensitivity (see \cite{AbbottProspects2018}), and a single 3rd generation detector with 10 times the horizon distance as LIGO at design sensitivity. The range shown for LIGO/Virgo's late phase indicates the uncertainty in expected sensitivity.}
\end{table}

We then calculated the detection volume for the joint observation of a GW and a GRB from the same source. Here, we calculate these volumes for the same commissioning phases of GW detectors as above. For simplicity we did not account for any sensitivity improvement due to combining GW and gamma-ray information. Results are shown in Table \ref{table:gw-only}.

While the GW signal from a BNS merger is essentially identical from all mergers, the gamma-ray output can vary orders of magnitude. We therefore additionally calculated detection volumes assuming different GRB isotropic-equivalent energies. We find that detection volumes vary only $O(20\%)$ in the range $E_{\rm \gamma,iso}=10^{50}-10^{52}$\,erg. This is due to the strong angular dependence of the gamma fluence; a large brightness change will correspond to a relatively small change in the threshold angle within which the GRB can still be detected.  Below we focus on results for $E_{\rm \gamma,iso}=10^{52}$\,erg, in line with the results of Margutti {\it et al.} \cite{2018ApJ...856L..18M}.

	We determined the fraction of BNS mergers discovered through GW emission that will also have a detected GRB counterpart by comparing the obtained GW-only detection volumes to the obtained GW+GRB detection volumes, both in (Table \ref{table:gw-only}) . We find that $\approx 35\%$ of GW detections will be accompanied by a GRB observation for both late and design GW phases, while this fraction is 27\% for a 3rd Gen. detector. For a fiducial top-hat GRB with $\theta_{\rm j}=10^\circ$, not considering Doppler effects, the corresponding fraction would be $\sim 2$\%.

\subsection{What fraction of observed short GRBs occurred in the local Universe?}

\cite{2005Natur.438..991T} identified a directional correlation between short GRBs and local galaxies, concluding that $10\%-25\%$ of short GRBs are likely to have been observed from within $100\,$Mpc. Considering that Fermi-GBM detects $\sim40$ short GRBs annually \cite{2016ApJS..223...28N}, this means that $4-10$ GRBs should be observed every year from within 100\,Mpc.  

For comparison, from the observed rate and distribution of distant GRBs, the local observed short-GRB rate density has been inferred to be $\sim10$\,Gpc$^{-3}$yr$^{-1}$ \cite{2006ApJ...650..281N}, corresponding to GRB observations within 100\,Mpc only once every 30 years. 

Structured gamma-ray emission can partially alleviate this discrepancy. More distant GRBs are only seen within their narrow energetic cone, resulting in an inferred lower rate, while local GRBs can be detected even from larger angles where weaker emission is compensated by the close distance. The weak structured emission also explains why observed nearby GRBs are not much brighter than distant ones, which concerned \cite{2005Natur.438..991T}.

To quantify this effect, we assume that short GRBs like GRB170817A are expected to be detectable within 100\,Mpc out to $\sim40^\circ$, while distant GRBs can only be detected out to their typical beaming angle of $\sim10^\circ$ \cite{2014ARA&A..52...43B}. This means that local short GRBs are detected at a rate $(1-\cos(40^\circ))/(1-\cos(10^\circ))\approx 15$ times higher than distant GRBs, corresponding to a local observed rate of $\sim150$\,Gpc$^{-3}$yr$^{-1}$. Based on this estimate, about 1\% of observed short GRBs should be located within $100$\,Mpc. This is significantly lower than the $10\%-25\%$ found by \cite{2005Natur.438..991T}. Nevertheless it means that there should be multiple short GRBs that have been detected from the local Universe with Fermi-GBM and Swift-BAT, without their distance having been identified. 

Based on the simulated structured jet profile we also find that $\sim10\%$ of observed short GRBs should occur within 200\,Mpc. The fact that such nearby GRBs have not been identified can be an observational selection effect: nearby events are typically detected off-axis, whose distances are difficult to measure. We will discuss this more in detail in a subsequent work.

\subsection{Constraints on BNS merger rate}

Using the simulated structured jet profile of Margutti {\it et al.} \cite{2018ApJ...856L..18M}, we find a GRB detection volume of $\sim0.1$\,Gpc$^{-3}$. Considering that Fermi-GBM and Swift-BAT observe about 45 short GRBs annually, the corresponding BNS merger rate is $\sim500$\,Gpc$^{-3}$yr$^{-1}$. This is consistent with the rate $320-4740$\,Gpc$^{-3}$yr$^{-1}$ obtained by LIGO/Virgo from the detection of BNS merger GW170817 \cite{2017PhRvL.119p1101A}, but higher than recent population synthesis estimates that put the rate in the range of $60-330$\,Gpc$^{-3}$yr$^{-1}$ \cite{Vangioni}. 

This rate is lower than the estimated rate of $\sim1000$\,Gpc$^{-3}$yr$^{-1}$ assuming highly beamed emission \cite{2014ARA&A..52...43B}. This is expected as structured emission increases the detectability of nearby events, therefore a fixed number of GRB detections corresponds to a lower BNS merger rate.

\section{Conclusion}

We investigated the observational consequences of structured $\gamma$-ray emission in short GRBs produced by BNS mergers. Structured jets lead to increased detectability of short GRBs in the local Universe ($\lesssim 200$\,Mpc) than for more distant sources due to weak emission out to larger viewing angles. We find the following observational consequences of this effect, using GRB170817 as a fiducial short GRB:
\begin{itemize}
\item A short GRB will be observed from more than 30\% of BNS mergers discovered via GWs, making such multi-messenger detections common.
\item About 10\% of observed short GRBs occurred within 200\,Mpc from Earth. This means that a significant fraction of short GRBs with no reconstructed distance are nearby.
\item The local rate density of BNS mergers is about $500$\,Gpc$^{-3}$yr$^{^-1}$.
\end{itemize}

These results assume that all short GRBs from BNS mergers are like GRB170817A, which is not necessarily the case. Future multi-messenger observations of BNS mergers will help improve these estimates. In addition, our results rely on the simulations of Margutti et al. \cite{2018ApJ...856L..18M}, which may be improved as more observations of the afterglow of GRB170817A, and more detailed numerical studies of the outflow, become available. Taking into account the changing $\gamma$-ray spectrum as a function of viewing angle will further improve the accuracy of the results. The structured jet profile in Margutti et al. \cite{2018ApJ...856L..18M} is provided only out to $40^\circ$. Presumably nearby GRBs may be detectable at even larger viewing angles, which makes our results, in this regard, conservative. 

Finally, there are also alternative emission models to consider, in particular the presence of a mildly relativistic, wide-angle outflow \cite{2017Sci...358.1559K,2018Natur.554..207M,2018ApJ...858L..15D,2018arXiv180609693M}, which will likely yield different predictions to the parameters obtained in this work. Near future observational constraints on these parameters will also provide strong constraints on structured emission.
\newline

The authors thank Szabolcs Marka and Daichi Tsuna for their useful suggestions. IB is thankful for the generous support of the University of Florida. This paper was approved for publication by the LIGO Scientific Collaboration and Virgo Collaboration under document number LIGO-P1800252.

\end{document}